\documentclass{article}

\usepackage{microtype}
\usepackage{graphicx}
\usepackage{subcaption}
\usepackage{booktabs} 
\usepackage[table]{xcolor}
\usepackage{hyperref}

\usepackage[preprint]{icml2026}

\usepackage{amsmath}
\usepackage{amssymb}
\usepackage{mathtools}
\usepackage{amsthm}
\usepackage[normalem]{ulem}

\usepackage[capitalize,noabbrev]{cleveref}

\theoremstyle{plain}

\theoremstyle{definition}

\theoremstyle{remark}

\usepackage[textsize=tiny]{todonotes}

\icmltitlerunning{CisTransCell: Single-Cell Perturbation Prediction via Gene Function, Regulatory Control, and Cellular Context}

\begin{document}

\twocolumn[
  \icmltitle{CisTransCell: Single-Cell Perturbation Prediction \\
  via Gene Function, Regulatory Control, and Cellular Context}



  \icmlsetsymbol{equal}{*}

  \begin{icmlauthorlist}
    \icmlauthor{Wei Zhang}{equal,comp}
    \icmlauthor{Xun Jiang}{equal,sch_1}
    \icmlauthor{Yuesi Xi}{yyy}
    \icmlauthor{Ming Tang}{comp}
    
  \end{icmlauthorlist}

  \icmlaffiliation{yyy}{Institute for Information Processing (tnt), Leibniz Universität Hannover, Germany}
  \icmlaffiliation{comp}{L3S Research Center, Leibniz Universität Hannover, Germany}
  \icmlaffiliation{sch_1}{School of Clinical Medicine \& Laboratory Medicine, Jiangsu University, China}

  \icmlcorrespondingauthor{Xun Jiang}{18627180758@163.com}

  \icmlkeywords{Machine Learning, ICML}

  \vskip 0.3in
]



\printAffiliationsAndNotice{}  

\begin{abstract}

Predicting cellular transcriptional responses to genetic perturbations is a central problem in single-cell biology, especially in the zero-shot setting where the perturbed gene or gene combination is unseen during training. 
A major difficulty is that perturbation effects are not determined by expression state alone: they depend on how the perturbed gene product influences other genes and proteins, how those downstream factors act on cis-regulatory elements, and which regulatory programs are active in the current cell state.
To better capture this biological complexity, we propose CisTransCell, a cell-conditioned multi-modal framework for single-cell perturbation prediction that augments each gene with two complementary priors: a regulatory-sequence prior that captures how the gene is controlled, and a coding-sequence prior that captures what the gene product does.
By integrating these priors with cellular expression state, CisTransCell models perturbation response as a cascade from gene function to regulatory control to downstream transcriptional change.
Experiments on benchmark single-cell perturbation datasets show that CisTransCell achieves strong performance in zero-shot perturbation prediction.

\end{abstract}

\begin{figure}[t]
    \centering
    \includegraphics[width=0.8\columnwidth]{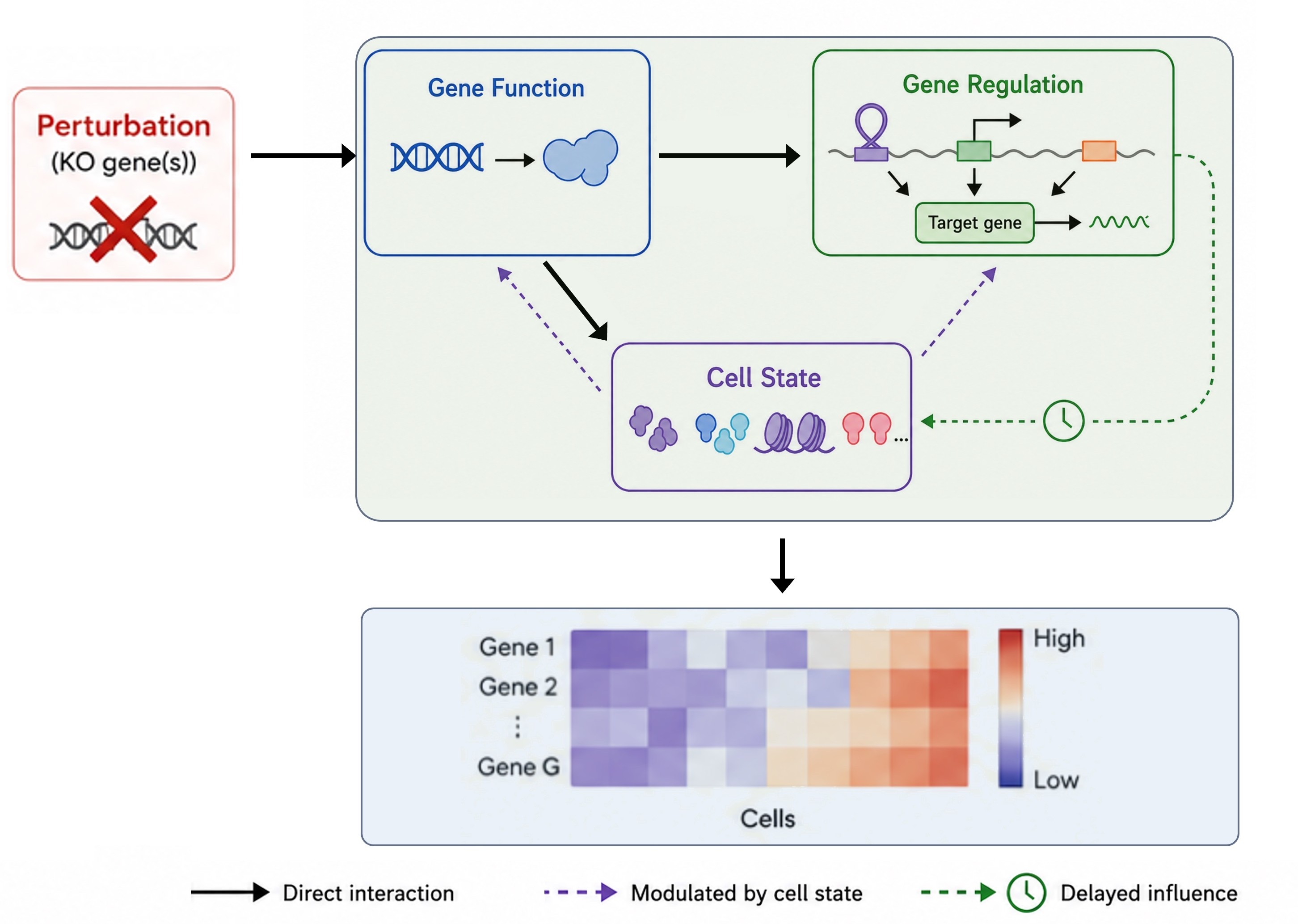}
    \caption{Perturbation response: perturbation propagates through three coupled components: gene function, gene regulation, and cell state. Gene function captures the downstream molecular influence of the perturbed gene product, gene regulation captures the cis-regulatory control of target genes, and cell state provides the cellular context that modulates both. Together, these components determine the post-perturbation transcriptional response.}
    \label{motivation}
\end{figure}

\section{Introduction}
A central goal of single-cell perturbation biology is to predict how genetic perturbations reshape cellular transcriptional states. Accurate \textcolor{black}{in silico} perturbation prediction could help prioritize experiments, improve mechanistic understanding, and accelerate functional genomics. This task is especially challenging in the zero-shot setting, where the perturbed gene or gene combination is unseen during training.

Existing methods for single-cell perturbation prediction often model perturbation effects in latent transcriptomic space or other learned representation spaces, including disentangled latent models, graph-based models, and foundation-model approaches \cite{cpa,gears,scgpt,scfoundation,state}. In these models, perturbed genes are typically represented through perturbation embeddings, graph nodes, or other learned features rather than explicit molecular priors. Although such representations can capture statistical regularities in the training data, they usually do not explicitly encode the molecular properties that govern how a perturbation propagates through the cell.

From a biological perspective, a gene participates in perturbation response through at least two distinct molecular roles. First, its regulatory-sequence determines how the gene is controlled, including which upstream factors can influence it and under what cellular conditions it can respond. Second, its coding-sequence determines what the gene product does, for example whether it functions as a transcriptional regulator, signaling component, enzyme, or structural factor. 
\textcolor{black}{As shown in \cref{motivation}, a perturbation propagates through a molecular cascade rather than a direct gene-to-expression mapping. The effect of knocking out a gene first depends on what the perturbed gene product normally does—that is, which genes, proteins, or pathways it can influence. Those affected downstream factors can then alter the activity of cis-regulatory elements, such as enhancers, promoters, insulators, and silencers, \textcolor{black}{which determine the transcriptional output of target genes.} Whether these regulatory effects are realized further depends on the current cell state, which specifies which regulators, pathways, and chromatin programs are active. Thus, perturbation response is jointly determined by gene function, regulatory control, and cellular context.}


\textcolor{black} { Motivated by this principle, we propose CisTransCell, a
cell-conditioned multi-modal framework that integrates genomic sequence priors with dynamic transcriptomic states. 
Our main contributions are: 
\begin{itemize}
    \item We propose to augment each gene with complementary regulatory- and coding-sequence priors, thereby providing a molecular context for each gene.
    \item We develop a biologically structured framework to model perturbation response.
    \item We demonstrate that CisTransCell achieves strong performance in zero-shot perturbation prediction across benchmark datasets.
\end{itemize}}

\section{Method}

\subsection{Problem formulation}

Let \(G\) denote the number of genes. Each training example is represented as a control-to-perturbation pair
\((x^{\mathrm{ctrl}}, y, p)\), where \(x^{\mathrm{ctrl}} \in \mathbb{R}^{G}\) is the expression profile of a control cell, \(y \in \mathbb{R}^{G}\) is the expression profile of a perturbed cell, and \(p\) denotes the perturbation target genes. To support both single-gene and two-gene perturbations in a unified form, we represent \(p\) as a length-two gene-index vector; for single-gene perturbations, the same index is repeated. The goal is to learn a perturbation-conditioned predictor:
\[
\hat{y} = f_{\theta}(x^{\mathrm{ctrl}}, p; R, C),
\]
where \(R \in \mathbb{R}^{G \times d_r}\) is a gene-level regulatory-sequence prior matrix derived from regulatory DNA sequence, and \(C \in \mathbb{R}^{G \times d_c}\) is a gene-level coding-sequence prior matrix derived from coding-sequence. Given a control cell and perturbation target(s), the model predicts the full post-perturbation expression profile \(\hat{y} \in \mathbb{R}^{G}\).

\subsection{Gene-level sequence priors}

Each gene is associated with two sequence-derived priors: a regulatory-sequence prior and a coding-sequence prior. Regulatory priors are derived from strand-aware genomic windows centered on the transcription start site, using a default window of 5500 bp upstream and 500 bp downstream. These sequences are embedded with DNABERT-2 \cite{dnabert} to produce one regulatory representation per gene.

Coding priors are derived from gene-associated transcript sequences. For each gene, we select the longest available cDNA transcript when possible, and otherwise fall back to an ncRNA sequence. Since transcript sequences can be long, we embed them with the Nucleotide Transformer \cite{nt} using overlapping chunks and combine the chunk-level representations into a single coding prior for each gene.

Both prior tables are aligned to the gene list of the dataset. Genes without a matched sequence prior are assigned zero-valued rows, preserving gene order and output dimensionality.

\subsection{Model architecture}

\begin{figure*}[t]
    \centering
    \includegraphics[width=0.75\textwidth]{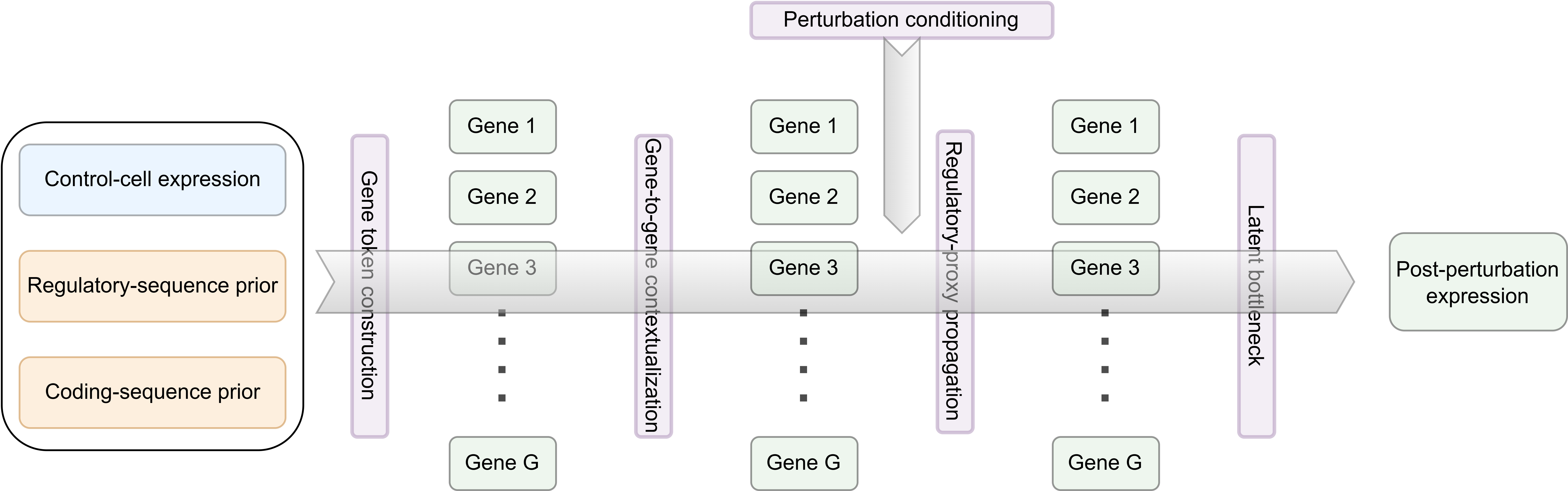}
    \caption{CisTransCell starts from the control-cell expression profile together with gene-level regulatory- and coding-sequence priors to construct one token per gene. These tokens then undergo gene-to-gene contextualization, perturbation conditioning, regulatory-proxy propagation, and a latent bottleneck, before being decoded into the predicted post-perturbation expression profile.}
    \label{model}
\end{figure*}

As shown in \cref{model}, CisTransCell combines the control-cell expression profile with gene-level regulatory and coding priors to construct one token per gene. These tokens then pass through gene-to-gene contextualization, perturbation conditioning, regulatory-proxy propagation, and a latent bottleneck, before a final decoder predicts the post-perturbation expression profile.

\subsubsection{Gene token construction}

CisTransCell combines static gene-level sequence priors with dynamic control-cell expression. For each gene \(g\), the regulatory prior \(R_g\) and coding prior \(C_g\) are first projected separately into modality-specific token spaces:
\[
\tilde{R}_g = \phi_r(R_g), \qquad
\tilde{C}_g = \phi_c(C_g),
\]
where \(\tilde{R}_g, \tilde{C}_g \in \mathbb{R}^{H}\) are the projected regulatory and coding representations. The two projected priors are then concatenated and fused to obtain a static gene token:
\[
S_g = \phi_f([\tilde{R}_g ; \tilde{C}_g]), \qquad S_g \in \mathbb{R}^{H},
\]
where \([\cdot;\cdot]\) denotes concatenation and \(\phi_f\) is the fusion MLP. Stacking all gene tokens yields a static gene token matrix \(S \in \mathbb{R}^{G \times H}\).

For a given control cell, we also construct dynamic per-gene features, including log-transformed expression, library-size-normalized expression, and a binary indicator of whether the gene is perturbed. These features are mapped into expression tokens \(E \in \mathbb{R}^{G \times H}\), and the initial gene representation is:
\[
Z_0 = E + S.
\]

This construction yields one hidden token per gene that jointly encodes sequence-level prior information and the current control-cell state.

\subsubsection{Gene-to-gene \textcolor{black}{(G2G)} contextualization}

The initial gene tokens are first contextualized through a self-attention mechanism, which allows each gene representation to incorporate information from all other genes:
\[
Z_1 = \operatorname{G2G}(Z_0),
\]
where \(\operatorname{G2G}\) denotes a self-attention block followed by a feed-forward update. This step captures gene-gene dependencies in the current cellular state before perturbation-specific conditioning.

\subsubsection{Perturbation conditioning}

Next, the model constructs a perturbation context from the target genes and uses it to condition all gene tokens. Let \(p\) denote the set of perturbed genes. We first aggregate the projected regulatory and coding priors of the perturbed genes:
\[
\bar{r}_p = \frac{1}{|p|}\sum_{k \in p} \tilde{R}_k,
\qquad
\bar{c}_p = \frac{1}{|p|}\sum_{k \in p} \tilde{C}_k,
\]
where \(\tilde{R}_k\) and \(\tilde{C}_k\) are the projected regulatory and coding representations of perturbed gene \(k\). These aggregated priors are then fused into a perturbation context vector \(h_p \in \mathbb{R}^{H}\):
\[
h_p = \phi_p([\bar{r}_p ; \bar{c}_p]),
\]
where \(\phi_p\) is a fusion MLP that maps the aggregated regulatory and coding priors into the hidden space.

The perturbation context is then applied to all gene tokens using feature-wise linear modulation (FiLM). To allow different genes to respond with different strengths, the model computes a gene-specific regulatory gate: 
\[
a_g = \sigma\left(\frac{(W_a \tilde{R}_g)^\top h_p}{\sqrt{H}}\right),
\]
where \(a_g \in (0,1)\) measures how strongly gene \(g\) is modulated by the perturbation context. The modulated representation of gene \(g\) is:

\[
\gamma_p = W_{\gamma}h_p,
\qquad
\beta_p = W_{\beta}h_p,
\]
\[
Z_{2,g} = Z_{1,g} \odot (1 + a_g \gamma_p) + a_g \beta_p,
\]
where \(\gamma_p, \beta_p \in \mathbb{R}^{H}\) are perturbation-specific scale and shift vectors, and \(Z_{1,g}\) is the contextualized gene token from the previous block, \(\odot\) denotes element-wise multiplication, and \(Z_{2,g}\) is the perturbation-conditioned gene representation. This mechanism allows the same perturbation to induce different responses across genes depending on their regulatory context.

\begin{table*}[t]
\centering
\caption{Perturbation Prediction Performance Comparison (in \%)}
\label{tab:prediction_performance}
\resizebox{\textwidth}{!}{
\begin{tabular}{lcccccccccccc}
\toprule
 & \multicolumn{4}{c}{Norman.} & \multicolumn{4}{c}{Rep1.} & \multicolumn{4}{c}{K562.} \\
\cmidrule(lr){2-5} \cmidrule(lr){6-9} \cmidrule(lr){10-13}
Models 
& $r_{\mathrm{pred,truth}} \uparrow$
& $r^{\mathrm{DEG}}_{\mathrm{pred,truth}} \uparrow$
& $\mathrm{ACC}_{\mathrm{pred,truth}} \uparrow$
& $\mathrm{ACC}^{\mathrm{DEG}}_{\mathrm{pred,truth}} \uparrow$
& $r_{\mathrm{pred,truth}} \uparrow$
& $r^{\mathrm{DEG}}_{\mathrm{pred,truth}} \uparrow$
& $\mathrm{ACC}_{\mathrm{pred,truth}} \uparrow$
& $\mathrm{ACC}^{\mathrm{DEG}}_{\mathrm{pred,truth}} \uparrow$
& $r_{\mathrm{pred,truth}} \uparrow$
& $r^{\mathrm{DEG}}_{\mathrm{pred,truth}} \uparrow$
& $\mathrm{ACC}_{\mathrm{pred,truth}} \uparrow$
& $\mathrm{ACC}^{\mathrm{DEG}}_{\mathrm{pred,truth}} \uparrow$ \\
\midrule
AverageKnown & 39.64 & 58.98 & 27.23 & 61.94 & 54.53 & 57.89 & 53.66 & 32.33 & 36.86 & 46.11 & 59.18 & 56.14 \\
Linear & 37.68 & 55.54 & 26.87 & 61.94 & 38.01 & 40.70 & 47.65 & 30.15 & 25.70 & 32.42 & 52.54 & 52.36 \\
Linear-scGPT & 39.20 & 58.66 & 27.23 & 61.94 & 50.09 & 53.95 & 49.69 & 31.31 & 33.86 & 42.97 & 54.79 & 54.37 \\
CellOracle & 9.80 & 12.48 & 19.20 & 16.35 & 39.91 & 7.40 & 37.55 & 23.70 & 4.44 & 5.89 & 41.41 & 41.15 \\
samsVAE & 12.48 & 32.05 & 37.42 & 49.63 & 12.59 & 36.45 & 33.04 & 25.08 & 8.51 & 29.03 & 36.44 & 43.55 \\
GraphVCI & 12.02 & 30.66 & 27.95 & 33.95 & 14.39 & 36.30 & 41.41 & 25.08 & 9.73 & 28.91 & 45.66 & 43.55 \\
scFoundation & 60.79 & 65.65 & 35.66 & 62.26 & 47.60 & 59.46 & 53.38 & 43.96 & 25.15 & 47.30 & 57.11 & 57.32 \\
scGPT & 61.48 & 65.87 & 61.96 & 74.43 & 50.32 & 65.54 & 61.72 & 67.07 & 32.72 & 43.15 & 57.44 & 57.32 \\
GEARS-Ens-10\% & 50.91 & 67.42 & 29.86 & 71.43 & 48.05 & 50.07 & 51.99 & 34.06 & 31.29 & 43.84 & 56.42 & 56.91 \\
PRESCRIBE-10\% & 64.32 & 68.61 & 64.73 & 75.93 & 60.28 & 66.13 & 67.89 & 80.03 & 38.58 & 47.52 & 61.04 & 71.21 \\
\rowcolor{gray!15} 
CisTransCell & 82.14 & 85.18 & 64.74 & 92.19 & 59.98 & 67.30 & 64.55 & 70.34 & 32.45 & 49.98 & 58.93 & 67.32 \\
\bottomrule
\end{tabular}
}
\end{table*}

\subsubsection{Regulatory-proxy propagation}

To model structured propagation of perturbation effects, CisTransCell introduces a smaller set of learned regulatory proxy tokens. These proxies serve as trainable latent representations of regulatory programs. Let \(A^{(s)} \in \mathbb{R}^{K \times H}\) denote the proxy-token matrix at propagation step \(s\), where \(K\) is the number of proxy tokens, and let \(Z^{(s)} \in \mathbb{R}^{G \times H}\) denote the current gene-token matrix.

For each gene \(g\), the model first computes a coding-prior-derived weight
\[
c_g = \sigma(\phi_{\mathrm{tf}}(\tilde{C}_g)),
\]
where \(\tilde{C}_g\) is the projected coding representation of gene \(g\), \(\phi_{\mathrm{tf}}\) is a learned scoring function, and \(c_g \in (0,1)\) controls how strongly gene \(g\) contributes to regulatory-proxy updates. Let \(c \in \mathbb{R}^{G}\) denote the vector of all gene-wise weights.

At propagation step \(s\), the regulatory proxy tokens \(A^{(s)}\) first attend to the gene tokens \(Z^{(s)}\), with the value stream gated by the coding-derived gene weights:
\[
A^{(s+1)}
= A^{(s)} +
\operatorname{Attn}\!\left(A^{(s)}, Z^{(s)}, c \odot Z^{(s)}\right),
\]
followed by a feed-forward block with a residual connection. The updated proxy tokens are then read back by the gene tokens,
\[
Z^{(s+1)}
= Z^{(s)} +
\operatorname{Attn}\!\left(Z^{(s)}, A^{(s+1)}, A^{(s+1)}\right),
\]
again followed by a feed-forward block with a residual connection.

The implementation repeats this two-step gene-to-proxy-to-gene exchange twice:
\[
Z^{(0)}
\rightarrow A^{(1)}
\rightarrow Z^{(1)}
\rightarrow A^{(2)}
\rightarrow Z^{(2)}.
\]
The output remains one token per gene, but the communication path now includes a learned regulatory-program bottleneck. We denote the final output of this block by \(Z_3\).





\subsubsection{Latent bottleneck}

After regulatory-proxy propagation, the gene-token matrix \(Z_3 \in \mathbb{R}^{G \times H}\) is refined with a Perceiver-style latent bottleneck. A learned latent set \(\Lambda \in \mathbb{R}^{M \times H}\) first attends to the gene tokens, undergoes two self-attention updates, and is then read back by the genes:
\[
B_0 = \operatorname{Attn}(\Lambda, Z_3, Z_3), \qquad
B_2 = \operatorname{SelfAttn}^{(2)}(B_0)
\]
\[
Z_4 = \operatorname{Attn}(Z_3, B_2, B_2).
\]

The refined gene tokens are decoded into a gene-wise expression change:
\[
\Delta_g = \phi_o(Z_{4,g}),
\]
and the final perturbed expression profile is predicted residually as:
\[
\hat{y} = x^{\mathrm{ctrl}} + \Delta.
\]

\subsection{Training objective}

Training data consists of paired control and perturbed cells. For each perturbed cell, a control cell is paired to form an example \((x^{\mathrm{ctrl}}, y, p)\). The model is optimized with AdamW using a loss that combines full-profile reconstruction with a perturbation-delta objective:
\[
\mathcal{L}
= \operatorname{MSE}(\hat{y}, y)
+ \lambda_{\Delta}\operatorname{MSE}(\hat{y} - x^{\mathrm{ctrl}}, y - x^{\mathrm{ctrl}}),
\]
where \(\lambda_{\Delta}\) is the weight of the delta-from-control term. The first term supervises the predicted perturbed profile, while the second term encourages accurate prediction of perturbation-induced expression change relative to the control state. We use the default setting \(\lambda_{\Delta}=0.5\).

\section{Experiment}
We follow the same experimental setup and evaluation protocol as \textsc{PRESCRIBE} \cite{cheng2025prescribe}, a recent state-of-the-art uncertainty-aware method for single-cell perturbation prediction published at NeurIPS 2025, so that our results are directly comparable to those reported in Table~2. Specifically, we evaluate on the Norman, Rep1, and K562 benchmark datasets, and report the same four prediction metrics:
$r_{\mathrm{pred,truth}}$,
$r^{\mathrm{DEG}}_{\mathrm{pred,truth}}$,
$\mathrm{ACC}_{\mathrm{pred,truth}}$, and
$\mathrm{ACC}^{\mathrm{DEG}}_{\mathrm{pred,truth}}$.

\cref{tab:prediction_performance} compares CisTransCell against the strongest reported variant in the \textsc{PRESCRIBE} paper, namely \textsc{PRESCRIBE}-10\%. Under the same benchmark setting, CisTransCell shows its clearest advantage on the Norman dataset, where it outperforms \textsc{PRESCRIBE}-10\% on all four metrics. On Rep1 and K562, CisTransCell remains competitive overall and achieves the best DEG correlation. These results highlight the advantage of CisTransCell in capturing structured downstream transcriptional effects induced by perturbation.

\section{Conclusion}

We propose CisTransCell, a biologically motivated framework for single-cell perturbation prediction. By unifying gene function, regulatory control, and cellular context, CisTransCell models how perturbation effects propagate from perturbed genes to downstream transcriptional responses. Across benchmark datasets, the model achieves strong overall performance. These findings underscore the value of biologically grounded priors for improving zero-shot perturbation prediction. In future work, integrating genomic priors with epigenomic context may further improve both accuracy and interpretability, and move toward more realistic models of how perturbations reshape cellular state.

\bibliography{example_paper}
\bibliographystyle{icml2026}

\end{document}